\documentclass[12pt]{iopart}

\usepackage[latin9]{inputenc}
\usepackage{graphicx}
\usepackage{textcomp}

\begin{document}

\title[Atomic spin coherence beyond the Johnson noise limit]{Cold atoms near superconductors: Atomic spin coherence beyond the Johnson noise limit}
\author{B Kasch$^1$, H Hattermann$^1$, D Cano$^1$, T E Judd$^1$, S Scheel$^2$, C Zimmermann$^1$, R Kleiner$^1$, D Koelle$^1$ and J Fort\'{a}gh$^1$}
\address{$^1$ Physikalisches Institut, Eberhard-Karls-Universit\"at
T\"ubingen, CQ Center for Collective Quantum Phenomena and their
Applications, Auf der Morgenstelle 14, D-72076 T\"ubingen,
Germany}
\address{$^2$ Quantum Optics and Laser Science, Blackett Laboratory,
 Imperial College London, Prince Consort Road, London SW7 2AZ, UK}
\ead{kasch@pit.physik.uni-tuebingen.de}

\begin{abstract}
We report on the measurement of atomic spin coherence near the surface of a superconducting niobium wire. As compared to normal conducting metal surfaces, the atomic spin coherence is maintained for time periods beyond the Johnson noise limit. The result provides experimental evidence that magnetic near field noise near the superconductor is strongly suppressed. Such long atomic spin coherence times near superconductors open the way towards the development of coherently coupled cold atom / solid state hybrid quantum systems with potential applications in quantum information processing and precision force sensing.
\end{abstract}

\submitto{\NJP}

\maketitle

\section{Introduction}
The construction of hybrid quantum systems which combine ultra-cold atoms with solid state devices has attracted considerable interest in the last few years. A particular goal has been the exchange of quantum information between gaseous atoms and microchip devices. One of the most promising proposals for achieving this involves coupling a Cooper pair box to a cold atomic cloud via a superconducting stripline cavity \cite{rabl_hybrid_2006, petrosyan_quantum_2008, petrosyan_reversible_2009}. Coupling of the Cooper pair box to the stripline cavity has already been achieved \cite{wallraff_2004} but the device will also require long-lived coherent coupling between cold atoms and the superconducting microwave cavity which has yet to be demonstrated.
To fully realize the potential of such devices, (e.g. the application as hybrid quantum processor, where gate operations are performed by the solid state quantum device and the atomic spin states serve as quantum memory) it will be necessary to bring the atoms within a few microns of the chip elements. However, the small separation between the atom cloud and the fabricated surface sets constraints on the atomic spin coherence.

Atoms trapped in a magnetic trap are subject to decoherence due to magnetic field fluctuations at the Larmor frequency, $\omega_L$, which change the spin state of the atom. A natural source of magnetic field fluctuations is the thermally excited motion of electrons in a metal (Johnson noise). Recent experiments have shown that the spin decoherence rate of atoms increases strongly when the atom cloud is trapped close to a conducting surface \cite{jones_spin_2003,harber_thermally_2003,lin_impact_2004,emmert_measurement_2009}. The measured decoherence rates are on the order of $1\,$s$^{-1}$ at a few tens of microns from bulk metals \cite{jones_spin_2003,harber_thermally_2003}  and about $10\,$s$^{-1}$ near room temperature metallic thin films at micron atom-surface separations \cite{lin_impact_2004}. The decoherence rate decreases when the surface is cooled, as demonstrated on a 4.2\,K gold thin film where rates of ~$0.1\,$s$^{-1}$ have been measured for distances down to $20 \,$\textmu m \cite{emmert_measurement_2009}.
The data confirmed theoretical predictions that Johnson-noise induced field fluctuations limit the spin coherence of atoms \cite{henkel_loss_1999, scheel_atomic_2005}.

Unlike normal conductors, superconductors are expected to shield magnetic field fluctuations. Atomic spin decoherence near superconductors has been predicted to drop well below the Johnson-noise limit of normal conductors \cite{hohenester_spin-flip_2007} and any other experimentally relevant loss rates, such as vacuum background collisions and tunneling of atoms from the trap to the surface \cite{lin_impact_2004}.

Here we report the observation of spin decoherence rates significantly below the Johnson noise limit in a hybrid system consisting of ultra-cold atoms and a superconducting surface. The coherence times are the longest yet observed in the vicinity of a highly conducting material and confirm the suppression of Johnson noise in superconductors. These results therefore support the suggestion that superconductors can be used to transfer quantum information between solids and gases.

\section{Cold atoms near superconductors}
\begin{figure}
\centerline{\scalebox{0.5}{\includegraphics{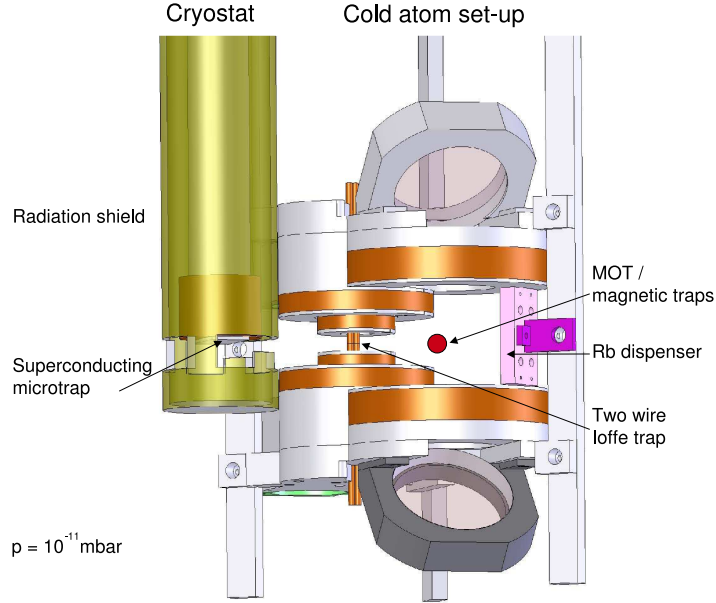}}}

\caption{In-vacuo trap set-up. The cold atom set-up (right) consists of a six-beam magneto-optical trap, magnetic quadrupole traps and a two-wire Ioffe-Pritchard type trap. The electromagnets are at room temperature. The MOT is loaded from a dispenser source. The $^4$He flow cryostat (left), is mounted next to the room temperature set-up. At the cold end of the cryostat we operate a superconducting magnetic microtrap for cold atoms. The background pressure is $p\approx 6 \times 10^{-11}\,$mbar.}

\label{interior}
\end{figure}

Our experimental system combines a cold atom set-up and a helium flow cryostat placed next to each other in ultrahigh vacuum at $6 \times 10^{-11}$\,mbar (figure \ref{interior}).
Initially, ultracold atoms are prepared via magneto-optical and magnetic trapping between room temperature electromagnets (right in figure \ref{interior}). The cryostat (left in figure \ref{interior}) holds superconducting structures on its cold end at a temperature of 4.2\,K and is protected against thermal radiation by a thermal shield at a temperature of 20\,K.
 
At the beginning of the experimental cycle, a six beam magneto-optical trap (MOT) is loaded from a pulsed rubidium dispenser. After standard polarization gradient cooling and optically pumping, the atoms in the $| F = 2, m_F = 2 >$   state are captured in a magnetic quadrupole trap and are subsequently transferred into a two-wire Ioffe-Pritchard type magnetic trap \cite{silber_2005}. The cloud is further cooled by forced radio frequency evaporation to a temperature of $2.5\,$\textmu K before loading it adiabatically into the dipole potential of an optical tweezers beam (1064\,nm, 0.8\,W, beam waist $30\,$ \textmu m). The optical tweezers are used for transporting the atoms to below the cryostat \cite{cano_meissner_2008}.
The horizontal translation of the beam waist over a distance of 44\,mm is accomplished within 570\,ms by translating the focussing lens of the tweezers beam outside the vacuum chamber on an air-bearing translation stage. At the cold end of the cryostat, we routinely load $10^6$  $^{87}$Rb atoms at a temperature of ~$1\,$\textmu K into a magnetic microtrap that is generated near a superconducting niobium wire ($d= 125 \,$\textmu m diameter). The radial confinement is defined by the magnetic field of the current carrying niobium wire and an external bias field perpendicular to the wire \cite{fortagh_magnetic_2007}.

 The niobium wire is clamped between solid copper blocks to ensure a good thermal contact to the cryostat (figure \ref{Angle_approach}). Due to the metallic contact between niobium and copper, the magnetic trap can only be operated with the niobium wire in the superconducting state. The axial confinement is given by a pair of parallel pinch wires separated by 2\,mm, running perpendicular to the niobium wire, mounted in the copper holder above the niobium wire \cite{cano_meissner_2008}.
 An external homogeneous field parallel to the niobium wire axis is used to control the offset field in the magnetic trap. The distance between atom cloud and niobium wire $d_{Nb}$ is controlled by the wire current (Fig \ref{Angle_approach}). The atom cloud can be moved in addition on a circular trajectory to the surface of the copper block by rotating the external bias field.
 
We calibrate the distance between atom cloud and niobium wire surface by means of in-situ absorption images. The position of the atoms is measured for various currents in the niobium wire. For small currents the distance between the trap and the wire surface is shortened by the Meissner effect, deviating strongly from an normal conducting trap, where the distance to the wire center is strictly linear to the current. By creating a fit to the  theoretical trap position (Eq.~(1) of ~\cite{cano_meissner_2008}), with the bias field as a free parameter and including gravity, we obtain the cloud-surface separation with an accuracy of $\pm2\,$\textmu m for any applied current.

The distance between the cloud and the copper surface is calibrated by imaging the circular trajectory on which the atom cloud moves when the bias field is rotated (figure \ref{Angle_approach}). The current in the niobium wire and the modulus of the bias field are kept constant, thus the angle of rotation is given by the ratio of the vertical and horizontal magnetic fields. The position of the surface is determined by measuring the fields where all atoms are immediately lost. The method gives an accuracy of $+1/-3\,$\textmu m.

\section{Exceptionally long atomic spin coherence near superconductors}
Magnetic traps allow for a rather simple method of measuring the spin coherence of an atomic ensemble. Since in a conservative magnetic potential only the low field seeking spin states are trapped, the spin decoherence rate of atoms can be derived from measurements of the magnetic trap lifetime \cite{jones_spin_2003,harber_thermally_2003,lin_impact_2004}. We apply this method for measuring the spin coherence near superconducting niobium.

\begin{figure}
\centerline{\scalebox{0.95}{\includegraphics{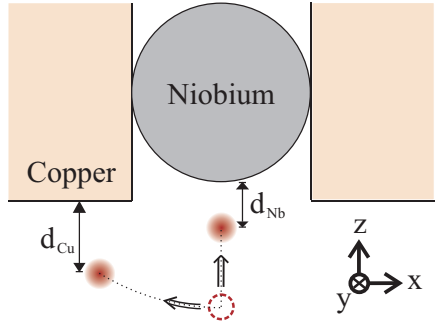}}}
\caption{Geometry of the experiment for measuring atomic spin decoherence near superconducting niobium and normal conducting copper at $T = 4.2\,$K. The 125 \textmu m niobium wire is clamped between solid copper in order to ensure thermal contact. From an initial position, the cloud is moved closer to the niobium wire by reducing the current in the wire, or closer to the copper surface by rotating the external bias field about the wire axis.   }
\label{Angle_approach}
\end{figure}

For our measurements, we use the following experimental sequence.  The $^{87}$Rb atom cloud is loaded into the magnetic microtrap near the superconducting niobium wire.
The trap has radial frequency $\omega_{r}=2\pi \times 138 \,$Hz, axial frequency $\omega_{a}=2\pi \times 21 \,$Hz. The Larmor frequency is $\omega_L=10.73\,$MHz for the spin polarized hyperfine state $|F = 2, m_F = 2 >$ at $B_{off}=2.44$\,G. The cloud is cooled by forced evaporation for 11\,s, whereby the temperature and atom number are reduced to $T \approx 100$\,nK and $N \approx 2 \times 10^5$, respectively. The atom cloud is then moved in 1\,s to a distance $d_{Nb}$ from the niobium surface (figure \ref{Angle_approach}) and held there for a variable hold time $t_{hold}$.  It is subsequently shifted away from the surface within 200\,ms after which the trap is turned off and the atoms are counted by absorption imaging after 5\,ms time-of-flight. 

\begin{figure}
\centerline{\scalebox{0.65}{\includegraphics{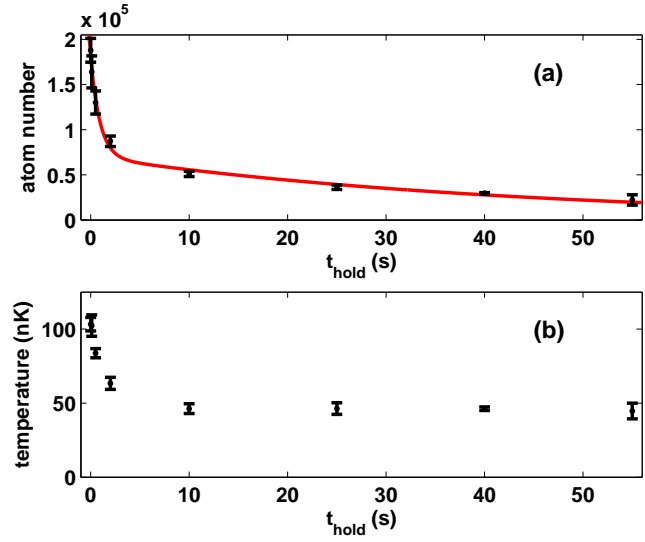}}}
\caption{Time evolution of number (a) and temperature (b) of an atom cloud held in a magnetic trap at a distance of $25\,$\textmu m from niobium.  
After an initial atom loss due to surface evaporation on time scale of $1/\Gamma_1$ (see text), the cloud decays exponentially with the rate $\Gamma_2$ and without measurable change in temperature.}

\label{Num_and_Temp_decay}
\end{figure}

The characteristic decay of the atom number in the superconducting microtrap as a function of the hold time is plotted in figure \ref{Num_and_Temp_decay}a. Bringing the thermal atom cloud close to the surface leads to a truncation of the Gaussian density profile. The subsequent surface evaporation and rethermalization \cite{harber_thermally_2003} reduces the number of atoms on a time scale of $1/\Gamma_1$ to typically $5\times 10^4$ and the temperature to 50\,nK (figure \ref{Num_and_Temp_decay}b). We fit the data to the function
\begin{equation}
N(t_{hold}) = (N_0\cdot\exp(-\Gamma_1 t_{hold}) + N_1) \cdot \exp(-\Gamma_2 t_{hold}),
\end{equation}
which resembles the experimental situation. $N_0+N_1$ is the initial atom number with $N_1$ the asymptotic atom number after surface evaporation. For hold times longer than $1/\Gamma_1$, the atom number decays exponentially with a rate $\Gamma_2$, which contains all experimentally relevant loss mechanisms, such as vacuum background collisions and spin decoherence.

\begin{figure}
\centerline{\scalebox{0.65}{\includegraphics{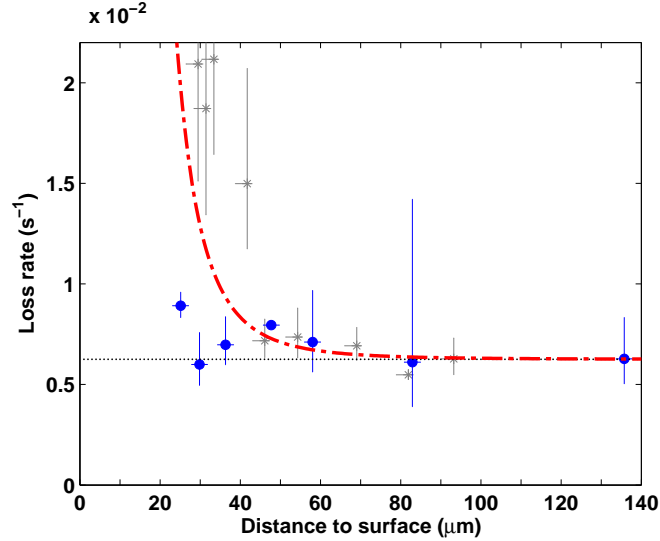}}}
\caption{Atom loss rate near superconducting niobium at 4.2\,K (blue circles) as a function of the atom-surface separation. The dash-dotted line (red) is the Johnson noise loss rate near normal conducting copper at 4.2\,K (see Eq. \ref{tau}) including the measured vacuum background loss (black dotted line). The data below the Johnson noise limit give evidence that magnetic field noise near superconductors is reduced compared with normal metals. A comparison measurement near copper is shown by the grey data points.}
\label{Both_rate}
\end{figure}

Figure \ref{Both_rate} summarizes the key results of this paper. We plot the measured loss rates $\Gamma_2$ near the superconducting niobium wire (blue circles) in order to compare them with the expected spin decoherence rate due to Johnson noise near a normal conducting metal (copper) at the same temperature.
The Johnson noise induced spin decoherence rate for copper is calculated from Eq. (4) in \cite{scheel_atomic_2005} by
\begin{equation}
\Gamma_{2} = \left(\frac{3}{8}\right)^{2}\frac{\bar{n}_{th}+1}{\tau_{0}}\left(\frac{c}{\omega_{L}}\right)^{3}\frac{3\delta_{L}}{d^{4}}
\label{tau}
\end{equation}
where $\delta_{L}=\sqrt{2/\omega_{L}\mu_0\sigma(T)}$ is the skin depth, $\sigma(T)$ is the electric conductivity of the metal, $\bar{n}_{th}\approx k_{B}T/\hbar \omega_{L}$ is the average occupation of thermal photons, $\tau_0 = 4 \times 10^{23} \,$s is the free space lifetime, and $d$ is the separation from the surface. This equation is valid provided that the skin depth is less than the atom-surface separation and also less than the metal thickness. The red dash-dotted line in figure \ref{Both_rate} is the calculated spin decoherence rate near copper at 4.2\,K ($\delta_L \approx 5\,$\textmu m), accounting for the lifetime limiting $| F = 2, m_F = 2 >$ to $| F = 2, m_F = 1 >$ transition, including an offset of $6.25 \times 10^{-3} \,$s$^{-1}$ that is the measured vacuum background loss rate.

The measured loss rates for atom-surface separations between $20\,$\textmu m and $40\,$\textmu m give clear evidence that the loss of atoms from the magnetic trap, and thus the spin decoherence rate near the superconductor, is below the Johnson noise limit of a normal metal. Magnetic field fluctuations near superconducting niobium are therefore strongly suppressed.

The diagram also shows a control measurement we made near a normal conducting copper surface at 4.2\,K (gray dots). The data confirm that the lifetime of the atom cloud is above the Johnson noise limit and agree with data reported elsewhere \cite{emmert_measurement_2009}. We conclude that near the copper (for the geometry of the measurement see figure \ref{Angle_approach}), Johnson noise and additional surface evaporation due to the limited trap depth are responsible for the observed loss rate.

For separations smaller than 20\,\textmu m between atom cloud and niobium wire, the Meissner effect reduces the magnetic trap depth \cite{cano_meissner_2008} and surface evaporation becomes the dominant loss mechanism. Nonetheless, this limitation of the actual microtrap configuration, which is caused by the exclusion of the magnetic flux from the niobium wire, can be overcome by using superconducting thin films \cite{cano_impact_2008} or superconducting microstructures with antidots \cite{kemmler_2006}. These allow the penetration of magnetic flux and promise to support strong magnetic confinement for atom clouds with exceptional spin coherence times even at micron distances from the superconducting surface.

\section{Conclusion}
Our measurements demonstrate that magnetic field fluctuations near a superconductor are suppressed below the Johnson noise limit of normal conductors. The loss rate we measured at 30\,\textmu m distance to the superconducting niobium wire is $6.25 \times 10^{-3} \,$s$^{-1}$, the smallest loss rate yet measured in the vicinity of highly conductive surface. These results are a step toward the coherent coupling of cold atoms with solid state systems and toward integrated precision force sensors, whereby cold atoms represent a quantum system of exceptionally long coherence time.

\section*{Acknowledgements}
This work was supported by the DFG (SFB TRR 21),
the UK EPSRC,
and the BMBF (NanoFutur 03X5506).


\end{document}